\documentstyle[11pt,aaspp4,psfig]{article}

\begin{document}

\title{Eclipsing Binaries in the OGLE Variable Star Catalog.~II.\\
Light Curves of the W~UMa-type Systems in Baade's Window}

\author{Slavek M. Rucinski\altaffilmark{1}\\
e-mail: {\it rucinski@astro.utoronto.ca\/}}

\affil{81 Longbow Square, Scarborough, Ontario M1W~2W6, Canada}
\altaffiltext{1}{Affiliated with the Department of Astronomy,
University of Toronto and Department of Physics and Astronomy,
York University}

\centerline{\today}

\begin{abstract}
Light curves of  the contact systems visible in the direction
of Baade's Window have been analyzed using the first coefficients 
of the Fourier representation. The results confirm that the
geometric contact between components is usually weak.
Systems showing significant differences in the depths of
eclipses are very rare in the volume-limited sample to 3 kpc: only
2 among 98 contact systems show the difference larger than 0.065 mag;
for most systems the difference is $< 0.04$ mag. If this relative
frequency of 1/50 is representative, then one among 
12,500 -- 15,000 Main Sequence F--K spectral-type stars is either a
semi-detached or poor-thermal-contact system. Below the orbital
period of 0.37 day, no systems with appreciable differences in
the eclipse depths have been discovered. 
Since large depth differences are expected to be associated with the
``broken-contact'' phase of the Thermal Relaxation Oscillations, this
phase must be very short for orbital periods above 0.37 day and possibly
entirely absent for
shorter periods. In the full sample, which is dominated
by intrinsically bright, distant, long-period systems, 
larger eclipse-depth differences are more common with about 9\% of
binaries showing this effect. 
Sizes of these differences correlate with
the sense of light-curve asymmetries (differing heights of maxima)
for systems with orbital periods longer than 0.4 day
suggesting an admixture of semi-detached systems with accretion
hot spots on cooler components. The 
light-curve amplitudes in the full 
sample as well as in its volume-limited sub-sample
are surprisingly small  and strongly suggest
a mass-ratio distribution steeply rising toward more dissimilar 
components. Since the sky-field sample is dominated by contact
binaries with
large amplitudes, it is suggested that a large fraction of low
mass-ratio systems remains to be discovered among bright stars in the
sky. For a mass-ratio distribution emphasizing low values, an approach
based on the statistics of the inner contact angles of totally-eclipsing
systems may offer a better means of determining this 
distribution than the statistics of the variability amplitudes.
\end{abstract}

\keywords{ stars: close binaries - stars: eclipsing binaries -- 
stars: variable stars}

\section{INTRODUCTION}

The first paper of this series (Rucinski \markcite{ruc97}1997 = R97)
demonstrated the potential of W~UMa-type systems as tracers of
the Galactic structure and population content. This paper should be
consulted for several details on the OGLE eclipsing-binary 
sample and on its analysis. R97 used almost
exclusively the time-independent data, such as the
orbital periods, maximum brightness magnitudes and colors, 
and upper limits to the interstellar reddening and extinction. The
present paper will concentrate on properties of contact binaries 
which are accessible through analysis of the light curves. 

The only information from light curves used in R97 was in the
preparation of an
automatically selected sample of contact systems observed
with relatively small errors and fulfilling shape criteria  for contact
configurations. The algorithm, 
based on Fourier decomposition of light curves for all 933 eclipsing
systems in nine OGLE fields in Baade's Window (BW), defined a much 
smaller sample of 388 contact binaries (the ``restricted-'' or R-sample)
than the original visual classification of the OGLE project 
which contained 604 systems (the O-sample). For
both samples, the additional constraint was that the period should be
shorter than one day and that the $V-I$ colors should be available. 
The current paper will present results mostly for the R-sample. 

For consistency with R97, the same data from the OGLE Catalog, Parts I
-- III are used in this paper (Udalski et al. \markcite{uda2}
\markcite{uda3}\markcite{uda4}1994, 1995a, 1995b). The new data for
BW9 -- BW11 (Udalski et al. \markcite{uda5}1996) have not been included
here also because the extinction/reddening data are not available for
these fields.

It has been shown in R97 that the distances of  the W~UMa systems in
the BW are distributed rather evenly in space all the
way to the Bulge. Determinations of distances required assumption of
the length scale of the interstellar absorption layer along the line
of sight, $d_0$. Two extreme assumptions on $d_0$ were considered:
(1)~extinction extending uniformly to the Bulge, $d_0 = 8$ kpc, and 
(2)~extinction truncated at $d_0 = 2$ kpc, with the latter assumption
giving a more consistent picture. The choice is however not critical
for the results in the present paper and enters only through a
definition of the local sample, judged to be complete to distance of 3
kpc. When applied to the R-sample, this choice
will be designated R$_2$, in consistence with R97.

The properties of the contact binaries which were extensively discussed
in R97 and will not be repeated here, although of
relevance for the present paper, were the following: Most
of the W~UMa-type systems 
belong to the population of old Turn-Off-Point stars. Their periods
and colors are confined to relatively narrow ranges of $0.25 < P <
0.7$ days and $0.4 < (V-I)_0 < 1.4$. The
{\it apparent\/} density of contact binaries 
is about $(7 - 10) \times 10^{-5}$
systems per pc$^3$ and the {\it apparent\/} 
frequency, relative to nearby dwarfs of similar colors
is one contact system per about 250 -- 300 
Main Sequence stars. A correction for undetected systems with low orbital
inclination would increase the above numbers by a factor of about 2 times.
Judging from these results, which were derived from 
the volume-limited sample, 
most of the contact binaries belong to the Old Disk population,
but with a possibility of an admixture of Halo and Thick Disk systems.

The present paper is organized as follows:
Sections \ref{fill} -- \ref{asym}
 contain discussions of various properties of the contact
binaries which can be analyzed using Fourier coefficients of the
light-curve decomposition. 
Section~\ref{fill} discusses the degree-of-contact (sometimes
also called ``over-contact'', although such an expression seems to be
awkward). Section~\ref{ptc} discusses the statistics of occurrence of
the poor thermal contact in contact binaries. Section~\ref{asym} 
addresses the poorly explored matter of light-curve
asymmetries. 
Sections~\ref{ampl} and \ref{total} address the matter of the
mass-ratio distribution determinations from 
the amplitude distribution and from the distribution of angles of
totality for totally eclipsing systems.
Section~\ref{sum} gives a summary of the paper.


\section{DEGREE OF CONTACT}
\label{fill}

As was discussed in R97, the two even cosine terms of the Fourier
decomposition of light curves, $a_2$ and $a_4$, can be
used to separate contact binaries from detached binaries (see
Figure~4 in R97). This
separation was used in the definition of the R-sample, which was
selected on the basis of (1)~a good overall fit of the cosine
representations to the light curves, with the mean standard 
error of the fit better than 0.04 mag (i.e. about 2-sigma
error for most systems), and (2)~a shape criterion based on the
$a_2$ and $a_4$ coefficients. This criterion can be directly linked to
to the degree-of-contact parameter used in Rucinski \markcite{ruc93} 
(1993 = R93). As was shown in R93, a crude
estimate of the degree of contact can be made by interpolation
in the $a_4 - a_2$ plane. Figure~1 shows
the O-sample and R-sample data, compared with the theoretical
predictions of R93. 
The three domains marked in the figure are defined as
envelopes for all combinations of mass-ratios ($0.05 \le q \le 1$) 
and inclinations ($30 \le i \le 90$ degrees) considered in R93,
 for three cases of the
degree of contact: when the stars just fill the inner critical
envelope ($f=0$, the steepest rising and the
narrowest domain), when the stars
fill the outermost common envelope ($f=1$, the largest and 
lowest domain in the figure) and for one intermediate
case ($f=0.5$). The upper edge of the marginal
contact domain is basically the same as the cut-off line for the 
Fourier filter: $a_4 = a_2\,(0.125-a_2)$.
The degree of contact  $f$ is defined in terms of the
potential, as in R93.

In addition to the R-sample, the O-sample is included in Figure~1
to show how many systems have 
been lost in moving from the original OGLE classification to
the restrictive R-sample. A large fraction of rejected
systems had poorly
defined light curves, but some very interesting objects, 
very close to contact, did not pass the Fourier filter.
 One example of particular importance is the
shortest-known period Main Sequence binary \#3.038 (P = 0.198 day) 
which consists of two very similar, strongly distorted, 
but detached, M-type dwarfs (a full light-curve 
synthesis solution by Maceroni \& Rucinski is in preparation). 
Since it was decided to concentrate on genuine contact systems,
the R-sample will be used from this point on. However, one should 
realize the limitations of the selection, which was based
on the Fourier coefficients calculated in R93 
for one effective temperature and one spectral band ($V$-filter). 
Since the contact-binary variability is dominated by geometrical changes
and weakly depends on the atmospheric properties, 
it can be argued that these
limitations should not strongly affect any {\it relations\/} between 
the coefficients (as opposed to their numerical values). However,
 this approach
cannot really replace full synthesis solutions of the light curves,
especially for derivation of the values of $f$.

As we can see in Figure~1, most contact systems
occupy a band corresponding to moderate
degrees of contact, of about $0 < f < 0.5$. This is in agreement
with the previous results (Lucy \markcite{luc73}1973;
Rucinski \markcite{ruc73}1973; Rucinski \markcite{ruc85}1985,
Fig.3.1.9). However, this method of estimating $f$ can 
be used only in a qualitative sense. In addition, it
entirely loses sensitivity for small amplitudes and
nothing can be said about cases with $|a_2| < 0.1$. 
In their majority, these will be small mass-ratio systems, 
as with the decrease of $q$, sizes of components become 
progressively more different leading to shallower eclipses.

Figure~1 
contains interesting contribution to  the matter of the
most common values of the mass-ratio. The broken lines join loci of
$a_2$ and $a_4$ combinations which can be reached for edge-on orbits
(inclination $i=90^\circ$) for fixed
values of the mass ratio, 0.1, 0.3, 0.5 and 1.0. The concentration of
observational points close to the origin, with strong fall-off toward
the upper right in Figure~1, strongly suggests that large (close
to unity) values of the 
mass-ratio are very infrequent. The mass-ratio distribution 
is apparently skewed with a preference for small values. 
We will return to the
important matter of the mass-ratio distribution in Sections
\ref{ampl} and \ref{total}. 


\section{POOR THERMAL CONTACT AND 
SEMI-DETACHED SYSTEMS}
\label{ptc}

Only the two even cosine coefficients, $a_2$ and $a_4$ of the
Fourier decomposition of the light curves have been used so far.
Now we will  consider the first odd term, $a_1$. 
Figure~2 shows the two first cosine coefficients, $a_1$
and $a_2$, for the full sample of 933 eclipsing systems in the 
Catalog, divided into groups according to the original
OGLE classification. This figure corresponds 
to Figure~4 in R97 which gave the $a_2$ -- $a_4$ 
dependence for the whole OGLE sample.
The filled circles are EW-type systems according to the OGLE
classification, i.e. the
contact binaries. Open circles mark EB systems. In the OGLE Catalog, 
all  EB systems have periods longer than one 
day\footnote{Many of the EB systems classified by OGLE
would formally fulfill our Fourier filter for inclusion in the
R-sample, but we considered only systems with $P < 1$ day.}. Crosses
mark all remaining systems with classes E, EA and E?. These systems
frequently have eclipses of unequal depths, so that values
of $a_1$ for them may differ substantially from zero. 
Indeed, as in Figure~4 in
R97, we see a clean separation between the 
contact systems, occupying a band within $-0.02 < a_1 < 0$
(or difference in the eclipse depths less than 0.04 mag), 
and other binaries which sometimes show strongly
negative values of $a_1$. For contact binaries with good 
energy exchange between components,
the $a_1$ term is expected to be
very small (R93), reflecting almost identical depths
of eclipses, in accordance with almost constant effective temperature
over the whole contact configuration. The equality of the effective
temperatures is a defining feature of the contact binaries
of the W~UMa-type; it was initially one of the most difficult
properties to explain and led to development of the successful
contact model by Lucy\markcite{luc68}(1968). 

A closer look at Figure~2 reveals that a certain fraction of 
systems classified by OGLE as EW or contact binaries
also appear to have negative values of $a_1$. These are the 
systems  of interest in this section.
There are two reasons why systems appearing as contact binaries
may show differences in eclipse depths: Some of them may 
be very close, semi-detached binaries (SD), and some may be in
contact, but with the energy exchange constricted or diminished
for some reason. We will call the latter poor-thermal-contact 
(PTC) systems. There exists a third reason why
systems in good geometrical contact may
show deviations from the Lucy model. This is the so-called 
``W-type syndrome'' related to slightly higher surface
brightnesses of the less-massive components. It 
seems to be limited to cool systems and may be related to 
their chromospheric activity. 
There have been several attempts to explain this light-curve
and a small
temperature (about 5\%) excess of the less-massive component 
seems to be preferred (for an exhaustive discussion of the effect
for the prototype case of W~UMa and for the literature, 
see Linnell \markcite{lin91a} 1991a,
\markcite{lin91b} 1991b, \markcite{lin91c} 1991c).  
The W-type syndrome is too small
to be addressed in this paper and actually must 
be hiding in the spread of points
within $-0.02 < a_1 < 0$ in Figure~2. We are interested here
in much larger effects produced by the SD or PTC causes.

It is very difficult to distinguish observationally between contact
systems with inhibited energy exchange and semi-detached systems
with components close to the inner critical Roche
lobe. Systems which show large differences in depths
of eclipses, yet appear to be in contact, are sometimes called the 
B-type systems for their $\beta$~Lyrae-type light curves. 
A relation between short-period semi-detached
systems and contact systems is expected to exist on the basis of
the theoretically predicted Thermal Relaxation Oscillations 
(Lucy \markcite{luc76}1976; Flannery \markcite{flan76}1976;
for the most recent review, see Eggleton \markcite{eggl} 1996). 
In the simplest version, the TRO cycles should
consist of two main phases, the good geometrical
and thermal contact state and the semi-detached state.
 In reality, switching between these two
may involve the PTC state. Relative durations of the 
contact and semi-detached branches of the TRO cycles
are expected to scale as some larger-than-one power of the mass-ratio:
with the contact stage lasting long and the semi-detached stage
quite brief. But nothing is known how long the PTC state could last.
Observations do not give us a clear picture,
mostly because of the poor statistics, spotty coverage of 
stellar parameter space and difficulties with separation
of  the SD and PTC cases.

After the work of  Lucy \& Wilson \markcite{luc79}(1979),
which explicitly addressed the question of systems in the 
``broken-contact'' phase, 
several short-period eclipsing systems with unequal depths
of the minima have been studied by Ka\l u\.zny and Hilditch 
with their respective collaborators  (Ka\l u\.zny \markcite{kal83}1983,
\markcite{kal86a}1986a, \markcite{kal86b}1986b, 
\markcite{kal86c}1986c; Hilditch et al. \markcite{hild84}1984,
\markcite{hild88}1988; Bell et al. \markcite{bell90}1990). 
These and other results have been discussed in terms of
the temperature difference between the components 
in the compilation of Lipari \& Sistero \markcite{ls88}(1988).
These investigations did not give an answer about the evolutionary
state of  such systems. Some of them seem to be genuine PTC 
contact systems with components of  unequal temperatures, 
some may be in the semi-detached state (with either the more- or
less-massive components filling their critical lobes),
mimicking contact systems. Usually, a full set of
photometric and spectroscopic data is needed in
individual cases. One property however is clear:
 Irrespectively what produces the large 
differences in depths of  eclipses,
such systems are very rare, even in spite of
certain advantage in discovery relative to normal contact systems.
We see them only among systems having orbital period longer
than a certain threshold value: the SD or PTC 
binaries do not occur among very cool, short-period
systems. The border line is currently at about 0.4 days, defined by
 the shortest-period PTC system known at this time,
W~Crv with $P = 0.388$ day (Odell\markcite {odell} 1996).

The OGLE data give us a first chance to look into statistics of the
occurrence of unequally deep eclipses. However, we should note an 
important limitation of results based solely on such photometric data:
Unless we see total eclipses, we do not know
which star, more or less massive, is eclipsed at each
minimum. In view of the surface-brightness deviations from the
contact model, this is a serious complication. Here, we follow
the conventional way of counting phases,
from the deeper minimum, i.e. from the eclipse of the hotter star.
A more meaningful convention, normally used in 
light-curve synthesis programs, would be with 
phases counted from the eclipse of the more-massive star. 

Figure~3
 shows the same data as in Figure~2, but for the R-sample.
Its sub-sample with the distances smaller than 3 kpc (the reddening
model with $d_0 = 2$ kpc, designated as R$_2$, see the Introduction), 
will be from now on called the
``local sample''. As was discussed in R97, this
distance delineates a volume-limited sample of contact systems in
the OGLE data. We can immediately see that contact systems with
large differences in depths of minima are all distant, hence
intrinsically bright. Among the 98 R-sample 
systems to 3 kpc, only 2 have
$a_1 < -0.03$ (difference in depth larger than 0.065 mag), 
so that the phenomenon of unequal eclipse depths is very rare and 
affects only some 2\% of the systems. For the full
R-sample extending all the way to the Bulge, 
with the same magnitude-difference threshold, we have 36 among
388 or 9\% of all systems. 
The two systems in the local R-sample with unequal
depths of minima are \#3.012 and \#6.005.
Their orbital periods are 0.370 and 0.698 days. The first system
sets a new short-period limit of the occurrence of the
phenomenon, replacing W~Crv with 0.388 day. The light curve
for \#3.012 is shown in Figure~4. The system \#6.005 
(discussed in the next section, its light curve is in 
Figure~11)
is the only nearby systems with large eclipse-depth difference and
relatively long orbital period. 

The results on $a_1$ for the R$_2$ sample are shown as an
orbital-period dependence in Figure~5 and as the intrinsic-color
dependence in Figure~6. As was discussed
in R97, the range of the observed intrinsic colors for most systems
is relatively narrow, due to their concentration in the
Turn-Off Point region for an old stellar population.
The distribution is additionally compressed on the red side
due to progressive elimination with distance of faint, red 
systems from the full, magnitude-limited R-sample.
As we already said, the local systems, with a relatively 
wider range of colors, show small values of $a_1$. This 
applies also to the reddest
binaries which we will discuss now in a 
small detour from the main subject. 
 
The three reddest systems in Figure~6, 
with the observed  colors $V-I > 2$ and the
intrinsic colors $(V-I)_0 > 1.5$ belong to the local sample with
distances of 600 -- 1400 parsecs. These small distances may be
erroneous if the colors are red due to some observational
problems rather than to genuinely low effective temperatures,
as very red colors lead to low intrinsic
luminosities in our absolute-magnitude calibration. 
The red colors are really unusual when compared with
the current short-period, red-color limit for the contact
binaries determined by CC~Com with $P=0.2207$ day, 
$B-V=1.24$, $V-I=1.39$ and the spectral type about $K5$ 
(Rucinski \markcite{ruc76}1976; Bradstreet \markcite{brad85}1985).
 All three systems, \#3.053, \#7.112 and \#8.072,
are quite unremarkable as far as
their light curves are concerned. Only \#8.072
has a short orbital period of 0.284, in some accord with the color,
 but even in this case the color
is well beyond what has been observed before for contact systems:
$V-I = 2.04$ and $(V-I)_0 = 1.77$, implying $(B-V)_0 \simeq 1.4$.
\#3.053 is unusual in that it has a moderately long period of
0.466 day in combination 
with a very red color: $V-I = 2.59$ and $(V-I)_0 = 2.31$. 
The corresponding $(B-V)_0 \simeq 1.5$ implies the
spectral type as late as M1 or M2. The light curves of \#3.053 and
\#8.072 have moderately large amplitudes of $\Delta I = 0.45$ and
0.54 so that blending of images with other, very red stars would be
difficult to postulate. Such blending may be the explanation for the low
amplitude of \#7.112 ($\Delta I = 0.17$), which is unusually red
($V-I=2.58$) for its orbital period of 0.590 day. All three systems
require further observations. We note that the possible error of
0.1 in the OGLE mean colors for red stars cannot be an explanation here,
as the stars are simply too red.

\section{LIGHT-CURVE ASYMMETRIES}
\label{asym}

Light-curve asymmetries are fairly common in contact binaries. The
difference in heights of minima is sometimes called the O'Connell
effect (for earlier references, see Linnell \markcite{lin82}1982
and Milone et al. \markcite{mil87}1987). 
There is no generally accepted interpretation of such asymmetries, but the
most obvious explanations would be in terms of stellar spots or some
streaming motions deflected by Coriolis forces. The starspot
explanation seems to work well in cool systems which are expected to
be very active. Little is known about causes of persistent
asymmetries for systems of spectral types earlier than F-type since
large magnetic spots would be difficult to imagine to exist on these
stars. Sometimes, the asymmetry is so large that it must be caused by
some streaming/accretion phenomena. Of particular importance here
is the short-period ($P=0.301$ day), late-type (K3V) detached system
V361~Lyr which shows a huge asymmetry 
(Ka\l u\.zny \markcite{kal90}1990, \markcite{ka91}1991;
Gray et al. \markcite{gray95}1995)
apparently due to an accretion process currently
taking place between components\footnote{To the author's
knowledge, no radial-velocity of this important system has been 
performed so that all inferences about it
are still based solely on photometric data.}.
The maximum after the deeper eclipse is higher in this case,
indicating in-fall on the cooler component as the most 
likely direction of the gas streaming.

No statistics are currently available as to how large are the
asymmetries and how often do they occur. Selection of 
sky-field objects for
individual observations is obviously highly biased. A large
sample, such as the OGLE sample, which can be subdivided into
magnitude- and volume-limited ones, is of great usefulness here.
Before presenting the results, the same warning as in Section~\ref{ptc}:
Our photometric sample suffers from the 
ambiguity in the  origin of phases by half of the orbital period. 
If the asymmetry phenomena are driven by the Coriolis force, 
they are expected to show signatures possibly related
to the relative masses of components eclipsed at 
each minimum. We do not have this information, but instead we 
have information about the relative effective
temperatures of components. These might correlate with the masses, but
this is not obvious that this must necessarily be the case.

The asymmetries have been analyzed in the simplest possible way by
inclusion of one sine term in the Fourier decomposition. No major
differences exist in the results for the O- and R-samples, so that
only the latter sample will be discussed here. Figure~7 
contains the
sine coefficients, $b_1$, plotted versus the largest cosine
coefficient $a_2$. The sign of
$b_1$ indicates which maximum is higher after the deeper eclipse. 
No obvious
correlation between both coefficients seems to be present. However,
the next Figure~8, with the dependence of $b_1$
on the orbital period, brings up an interesting property: While for the 
local, mostly short-period systems we see about equal numbers
of positive and negative values of $b_1$,
for systems with periods longer than about 0.4 day, the positive
asymmetries dominate. This effect is not very strong, but is
definitely present, as is shown in the histogram of the
$b_1$ coefficients for 302 systems of the R-sample with
$P > 0.4$ day in Figure~9.  The distribution is only mildly
asymmetric: the mean is 
slightly positive, $+0.0013$, and the skewness  is not significant 
($0.33 \pm 0.22$). However, if we treat the both branches as
independent distributions, the $\chi^2$ test gives probability 
of only 0.0047 that the differences in numbers in 8 bins of
$|b_1|$ could be due to random fluctuations.
If we eliminate the two bins close to the origin ($-0.002$ to $+0.002$)
which are most susceptible to random fluctuations,
then the probability that the two branches are different due to
a statistical fluctuation decreases to as little as 0.0026. 
Thus, the difference in the two sides of the distribution is
significant at the level of  0.995 to 0.997. We see no way how the
asymmetry in the distribution of  $b_1$ could be generated by 
ways of data reductions. It must be real. Apparently, for long-period
systems, incidence of  the elevated maximum after the deeper 
eclipse is  higher than the opposite case. 
  
The most obvious candidate for the cause of the asymmetry in the
distribution of $b_1$ would be the Coriolis force
acting on gas streams between components. The sense of the
deviations induced by the Coriolis force is such that -- if we see
hot spots due to accretion phenomena (as seems to be the 
case for V361~Lyr) -- the visibility of spots {\it before
the secondary minima\/} implies that the spots are located on
the cooler components. This would be exactly
what should happen for binaries in the semi-detached state
with more massive, hotter components losing mass for their
less massive companions. At this point, without spectroscopic
data, this is a hypothesis rather than a definite statement.

If some of the light-curve asymmetries are caused by mass-transfer phenomena
in semi-detached systems of similar type, then we could expect 
a correlation between the sense and size of the asymmetry and the
difference in the depths of eclipses. Such a correlation cannot be
very tight as it would reflect the percentage admixture of the
semi-detached systems among all systems showing the light curve asymmetries.  
The scatter plot for $a-1$ and $b_1$ is shown in Figure~10 where
some correlation between these quantities
is indeed visible. The formal linear (Pearson) correlation coefficient 
$-0.29 \pm 0.08$ indicates presence of a correlation with
a relatively high significance 
(probability that the observed value resulted from a correlation 
of independent distributions is at the level 0.001). 
The three systems with the largest deviations in Figure~10
($a_1 \le -0.05$ and $b_1 \ge 0.01$), and thus the best candidates to
be the semi-detached systems with mass-transfer are the following:
\#0.111, \#3.094 and \#8.133. Their periods
are relatively long, 0.639, 0.912 and 0.509, which coupled with their
blue colors gives the distances 4.5, 7.9 and 5.3 kpc (under the R$_2$
assumption). System \#3.094 was suggested in R97 to be in the 
Galactic Bulge.

A typical light curve for a system showing a large asymmetry is shown
in Figure~11. T
he system is \#6.005 with the orbital period of 0.698
day which was already mentioned in the previous section
as one of the two nearby systems with
a large difference in the depth of the eclipses. Thus, this system
directly shows the correlation between $a_1$ and $b_1$ and may
be taken as a prototype of the class.

The dependence of  the $b_1$ coefficient on the intrinsic color 
is shown in Figure~12. This plot seems
to indicate that for very cool systems, negative $b_1$ may be more
common, but because of the small number of such systems, the
significance to this indication is low.

\section{AMPLITUDE DISTRIBUTION}
\label{ampl}

Amplitudes of light variations of contact binaries contain an
important information on the mass-ratio ($q$) distribution: Large
amplitudes can be observed only for large mass-ratios, close to unity;
for small mass-ratios, only small amplitudes can be observed. This
simple relation is illustrated in Figure~13 for a few theoretical
distributions calculated on the basis of R93 for mass-ratios within
narrow ranges of $\Delta q = 0.1$ (every second interval in $q$).
 We see not only the obvious
dependence between the largest possible amplitudes and the mean value
of $q$, but also the effects of total eclipses producing similar
amplitudes for large fractions of systems within $q$-dependent
ranges of inclinations close to $i=90^\circ$. 
The distributions in Figure~13 cannot be obviously
observed as weighting by the
mass-ratio distribution $Q(q)$ occurs. Exactly this weighting
opens up a possibility of
observational determination of $Q(q)$ which would be of great
importance for our understanding of formation and evolution of contact
binary stars.

Application of the line of reasoning described above is not easy, even
for such a large sample as the one being analyzed now. First of all,
for an inverse problem of determination of $Q(q)$ from the amplitude
distribution, $A(a)$, very good statistics for {\it each bin\/}
of $A$ is needed. 
This is the main obstacle why our attempts at determination
of $Q(q)$ through a solution of the integral equation
$A(a) = \int Q(q) K(q,\,a)\,da$ have been so far unsuccessful 
and will not be discussed here.
The second problem is that the magnitude-limited sample is 
biased by the presence of distant, bright systems. As was shown in
Section~\ref{ptc}, these are systems with the highest frequency of
occurrence of unequally deep eclipses. For such systems, primary minima are
deeper and secondary minima are shallower than for good thermal
contact models leading to a corrupted statistics of the
amplitudes (although use of the $a_2$ coefficient could possibly help here).
If we eliminate those bright and distant systems, and utilize only
the systems of the local sample (which seems to give a fair representation of
most typical systems), the sample becomes 
too small for definite applications. 
Finally, the distribution of the third geometrical
parameter, the degree-of-contact, is not known but it does have some
influence on the amplitude distribution,
judging by the theoretical results, as in Figure~14.

Figure~14 gives the observed distributions $A(a)$, where $a = \Delta I$ 
in the OGLE data, as well as some additional theoretical
predictions. The latter are shown by lines for the flat 
mass-ratio distribution $Q(q)$ and for two cases of  the
degree-of-contact, $f=0$ and $f=0.5$ (upper panel), as well as for 
for a strongly falling distribution, $Q(q) = 1 - q$ with $f=0$ (lower panel).
These theoretical distributions show some sensitivity
to the assumptions on the shape of $Q(q)$ and on the value of
$f$, but look very different from the observed distributions.
The full R-sample (upper panel) and the local R$_2$
sample to 3 kpc (lower panel) are shown as histograms. They
indicate dominance of low-amplitude systems. Large amplitude
systems are almost non-existent in the OGLE sample, which
agrees very well with the old open-cluster data (Ka\l u\.zny \&
Rucinski \markcite{kr93}1993; Rucinski
\& Ka\l u\.zny  \markcite{rk94} 1994). The most obvious
explanation for dominance of small amplitudes would be by
a $Q(q)$ distribution which steeply increases for small values of $q$. 

The OGLE sample gives us a very different picture than for the whole
sky. In the sky field, we observe several bright 
contact systems which have large variability amplitudes, and some
of them indeed have spectroscopically-determined
mass-ratios very close to unity. The extreme cases
are SW~Lac with $q = 0.73 \pm 0.01$ (Hrivnak \markcite{hriv92}1992), 
OO~Aql  with $q=0.84 \pm 0.02$ (Hrivnak \markcite{hriv89}1989)
and VZ~Psc with $q=0.92 \pm 0.03$ (Hrivnak \& 
Milone \markcite{hriv89a} 1989). Thus, although $q=1$ 
apparently never happens,
some mass-ratios can be quite close to unity.
The large values of $q$ for the field systems are consistent with the
 amplitude distribution for the whole sky which
peaks at $a \simeq 0.55$ and extends to amplitudes
as large as one magnitude (Rucinski \& Ka\l u\.zny
\markcite{rk94}1994; shown as a 
dotted histogram in the upper panel of
Figure~14). Judging by the OGLE data, these 
are atypical cases, emphasized by the ease with which large-amplitude 
systems are detected in the sky surveys.
More typical contact binaries show small variation
amplitudes which are related to their small mass-ratios. Such systems
 remain to be discovered among bright stars of the sky field.

\section{MASS-RATIO DISTRIBUTION FROM 
TOTALLY ECLIPSING SYSTEM}
\label{total}

Since inversion of the amplitude distribution 
for derivation of $Q(q)$ presents several problems,
one can consider a different approach: Mochnacki \& 
Doughty \markcite{moch72a}\markcite{moch72b} 
(1972a, 1972b) described
a method of element determinations utilizing the angles of
internal eclipse contact for totally eclipsing systems. This
method  can obviously be applied only to a sub-sample of all systems,
but has an advantage of being 
very weakly sensitive to the degree-of-contact: Basically,
for contact configurations,
only two geometrical parameters, $q$ and $i$, determine the
the angle of the inner eclipse contact, $\phi = \phi (q,\,i)$. 

The method of determination of the mass-ratio
distribution $Q(q)$ would be quite simple:
The distributions $Q(q)$ and of the orbital
inclinations $I(i) = \sin i$ must
be statistically independent. Thus, from
the distribution of the inner-contact angles, $\Phi(\phi)$, one could
determine $Q(q)$ by solving the integral equation:
\begin{displaymath}
\Phi(\phi) = \int_{q(0^\circ,\phi)}^{q(90^\circ,\phi)}
             Q(q) \, \sin[i(\phi,q)]\>
             \vert \partial i(\phi,q)/\partial \phi \vert \> d q
\end{displaymath}
At present, we have not been able 
to apply this approach because the sample of totally
eclipsing systems among the OGLE systems was too small. Also,
because of the relatively large observational errors of about 0.02
mag, it was difficult to set up an automatic-selection process of
finding totally-eclipsing systems. However, the approach holds a
great potential and, in the future,
 should be used on large samples of well observed
systems. Since, as we have shown in Section~\ref{ampl},
 the mass-ratio distribution must be skewed to
small values of $q$, a relatively large fraction of contact binaries
should show total eclipses.

\section{SUMMARY}
\label{sum}

The paper contains analysis of light curves of the W~UMa-type
binaries in the OGLE Catalog of Periodic Variable Stars for
fields BWC -- BW9. It is a continuation of R97, but concentrates
on properties of the contact systems, rather than on their usefulness
for galactic-structure studies. The important result of R97
that the contact binaries belong mostly to the Turn-Off Point
population of old stars is not amplified here; the stress is
on structural properties of the systems, as they can be gauged
using simple methods of light curve characterization. Since
the observations have moderate accuracy of about 0.02 mag,
the low-order Fourier decomposition was judged to provide
an adequate tool for such a characterization.

Rough estimates of the degree-of-contact, obtained using
the $a_2$ and $a_4$ cosine coefficients confirm that the most
frequent values are concentrated with $0 < f < 0.5$, 
suggesting weak contact. The quality of this determination is
low and it  is somewhat qualitative. A much more interesting results
came from the analysis of  the differences in depths of eclipses
which was based on the first cosine coefficient, $a_1$. In
the volume-limited ``local'' sample to 3 kpc, among 98
systems, only 2 show appreciable depth differences, so that good
thermal and geometrical contact is a norm rather than an
exception. One of these systems, \#3.012, sets a new short-period
limit of 0.370 day for the occurrence of unequal eclipses.
For the full magnitude-limited
 sample, which is favorably biased toward the 
intrinsically bright, distant systems, incidence of the large 
depth differences is more common with some 9\% showing
differences in depths of minima larger than 0.065 mag ($a_1 < -0.03$).
Some of these may be in good geometrical
contact and  with the inhibited energy exchange, the
poor-thermal-contact systems (PTC), but some
may be actually semi-detached (SD) systems,
very close to the contact configuration. The present data, in only one
spectral band and without spectroscopic support, 
do not permit to distinguish between these possibilities
in individual cases. 

Analysis of the light curve maximum-light
asymmetries -- measured by the first sine term $b_1$ --
which weakly (but significantly) correlate with the
differences in the depths of the eclipses for periods
longer than about 0.4 day, suggests that the semi-detached
state with mass-exchange is common among systems
with unequally deep eclipses. The asymmetries
would be then caused by mass-exchange streams 
impinging outer layers of cooler components. Since the asymmetries
may be also caused by photospheric spots, the
statistics of asymmetries cannot be used to determine
the number of SD systems mimicking contact binaries. 
The second system in the local sample which shows a
large difference of the eclipse depths, \#6.005, 
is a perfect illustration of
 the correlation between the difference in the depth of eclipses 
and the sense of the light curve asymmetry.
The overall rarity of  systems with large
eclipse-depth differences suggests that the admixture
of  semi-detached or poor-thermal-contact
systems to the totality of contact systems is very small.
These are however the intrinsically brightest systems
which are visible to large distances so that they are preferentially
represented in magnitude-limited samples.
In the future, it would be highly desirable to obtain light curves
in a few spectral bands, say in the $U$, $B$, $V$ and $I$ filters. This
would permit decoupling of the effects of the accretion streams and
temperature differences between components from the geometric issues
of contact versus semi-detached configuration.

Presence of only two SD/PTC systems
among 98 contact binaries of the local sample
can explain why we do not see such systems among bright
stars. Using the apparent frequency of contact binaries found in R97
of 1/250 -- 1/300, we can estimate the
expected frequency of SD/PTC systems among Main Sequence dwarfs 
to be about 1/12,500 -- 1/15,000; since this estimate is based on 2
cases, it carries a Poissonian uncertainty of a factor of
about 1.4. A complete sample of bright stars to $V
\simeq 7 - 8$ is accessible from the 
Hipparcos Input Catalogue (CD-ROM Version, Turon et
al. \markcite{tur94}1994). By counting
luminosity-class stars IV and V in the HIP,
within the color range where contact
binaries occur (R97), $0.4 < B-V < 1.2$, to the successively deeper
limiting magnitudes $V =5$, 6, 7, and 8, 
one obtains 128, 497, 1620 and 4561 stars. 
Thus, no SD/PTC systems are expected to the brightness 
levels at which the Hipparcos sample starts showing selection
effects. However, by going one magnitude deeper, we could expect some
3--4 time more stars, so that we should be able to detect one or 
two SD/PTC system. This is well confirmed by the actual
numbers as the brightest among such systems, AG Vir and FT Lup, appear at
$V=8.5$ and 9.2 (Ka\l u\.zny \markcite{kal86b} \markcite{kal86c}
1986b, 1986c). As an aside, we mention here that the sample of bright
contact binaries, when compared with the Hipparcos numbers,
fully confirms the apparent frequency of 1/250 -- 1/300: 
To the same $V$-magnitude limits as above, the variable-star lists
contain 3, 3, 6, and 14 contact systems.

The distribution of the variability amplitudes suggests that the
mass-ratio distribution $Q(q)$
increases toward small values of $q$. Since
the increase appears for $\Delta I < 0.6$, but is modified by discovery
selection already below $\Delta I < 0.3$, the present distribution is
just too narrow to attempt a determination of the mass-ratio
distribution $Q(q)$. A distribution emphasizing low values of
$q$ should lead to frequent occurrence of totally eclipsing systems.
Such systems may allow an entirely independent determination 
of $Q(q)$ based on
the angles of the inner eclipse contacts. We note that the sky-field is
surprisingly devoid of low-amplitude contact systems; we suspect that
they have been simply overlooked in non-systematic surveys, a result
 which has led
to an over-representation of large-amplitude contact systems.

\acknowledgments
Thanks are due to Dr. Bohdan Paczy\'nski and Dr. Janusz Ka\l u\.zny
for useful comments and
suggestions and to the OGLE team for making their
data available through the computer networks.

Research grant from the Natural Sciences and Engineering 
Council of Canada is acknowledged here.

\bigskip
\noindent
Note: The data on the W~UMa-type stars in the OGLE survey of
fields BWC to BW8 which were used in R97 and in this paper are
available from the author or from 
http://www.astro.utoronto/$\sim$rucinski/rucinski.html

\newpage

\begin{figure}           
\centerline{\psfig{figure=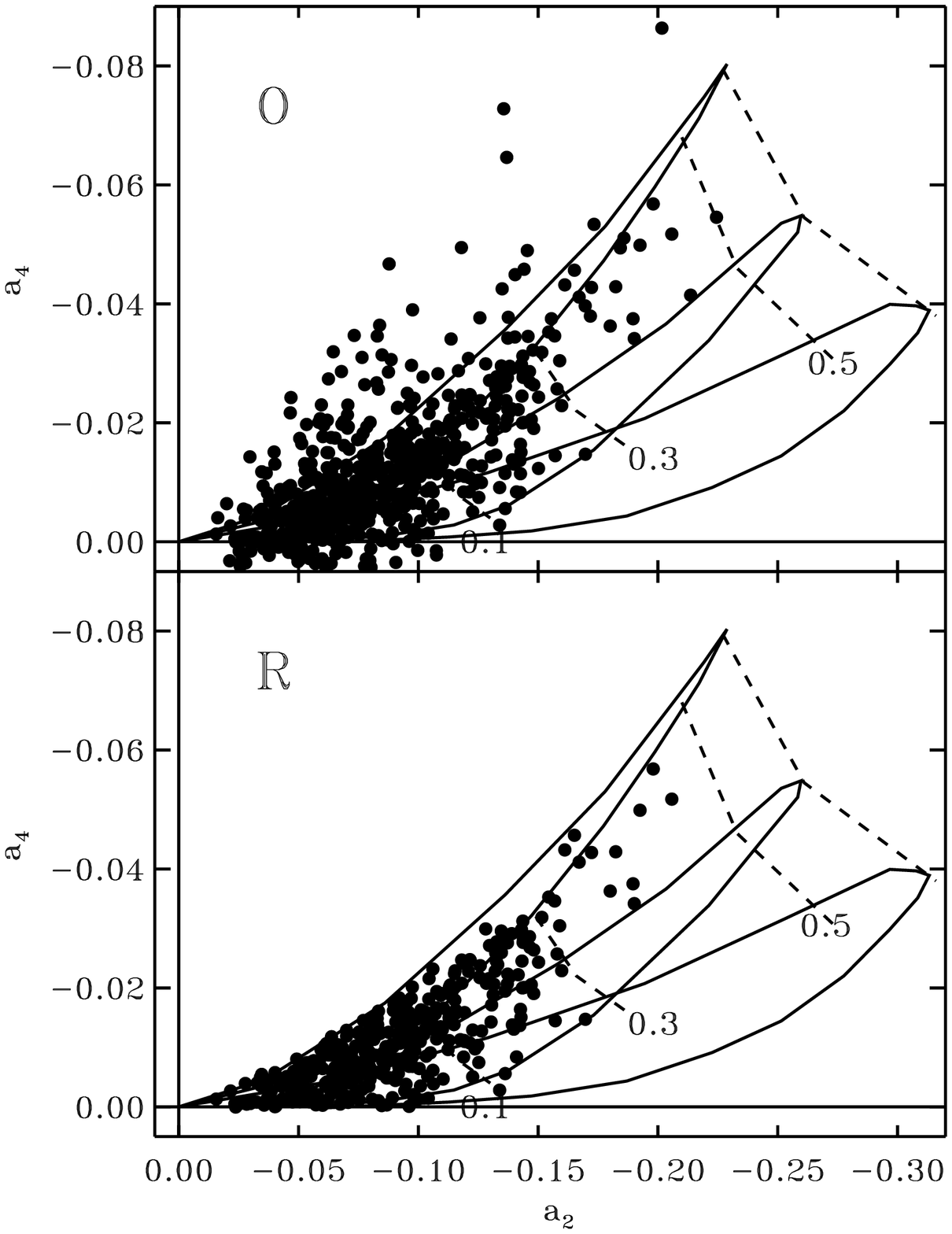,height=4.5in}}
\vskip 0.5in
\caption{Relations between the even Fourier coefficients $a_2$ and $a_4$
for the O-  and R-samples of contact binaries. The three elongated
regions give envelopes of all combinations of orbital inclinations and
mass-ratios considered in R93 for three values of  the
degree-of-contact parameter $f$. 
The upper, thin one is for the inner (marginal)
contact ($f=0$) and the lowest, broad one is for the outer contact
($f=1$); the intermediate is for $f=1/2$. The broken lines give
schematic locations of the 
combinations of the coefficients for $i=90^\circ$ for four values
of the mass-ratio $q$ of 0.1, 0.3, 0.5 and 1.0.}
\end{figure}

\begin{figure}           
\centerline{\psfig{figure=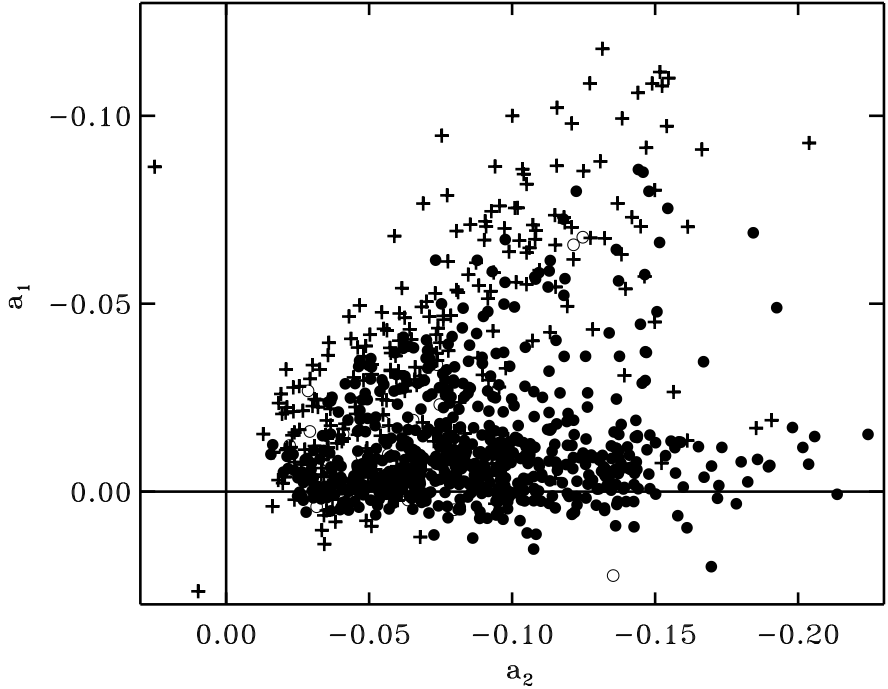,height=3.5in}}
\vskip 0.5in
\caption{Relations between the Fourier coefficients $a_1$ and $a_2$
for all
933 eclipsing systems as classified 
in the OGLE Catalog (fields BWC -- BW9, the
same as in R97). $a_1$ is sensitive to the difference in depth of
eclipses while $a_2$ is proportional to the overall amplitude of
light variations. The symbols are as follows: filled circles --
W~UMa-type systems (EW), open circles -- $\beta$~Lyrae-type systems (EB,
in the OGLE classification, all these systems 
have periods longer than one day), crosses -- all remaining 
eclipsing systems (E, EA, E?). Note that most contact systems
show small eclipse-depth differences within  $-0.02 < a_1 < 0$.
The group of systems showing larger differences may consist of genuine
contact systems with poor thermal contact (PTC) and of short-period
semi-detached (SD) systems mimicking contact systems.}
\end{figure}

\begin{figure}           
\centerline{\psfig{figure=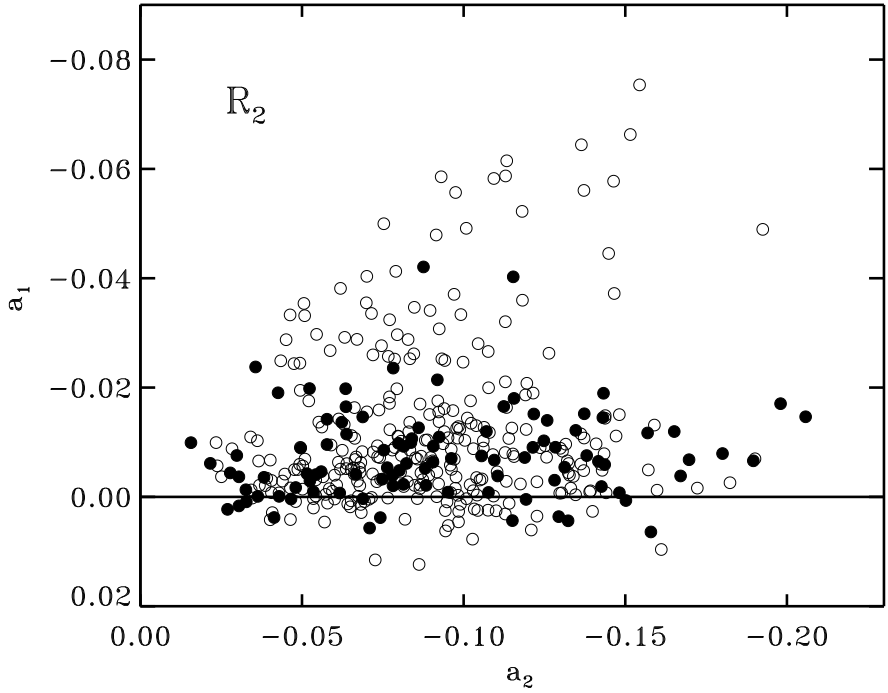,height=3.5in}}
\vskip 0.5in
\caption{The same plot as in Figure~2, but limited to the
contact systems
of the R-sample. In this and in subsequent figures,
systems forming the ``local sample'' to 3 kpc,
which we think is complete, are marked by filled symbols. Note
that PTC and SD systems are rare in space: In the local sample
consisting
of 98 systems, there are only two cases with $a_1 < -0.03$ or,
equivalently, with depths of eclipses differing by more than 0.065
mag.}
\end{figure}

\begin{figure}           
\centerline{\psfig{figure=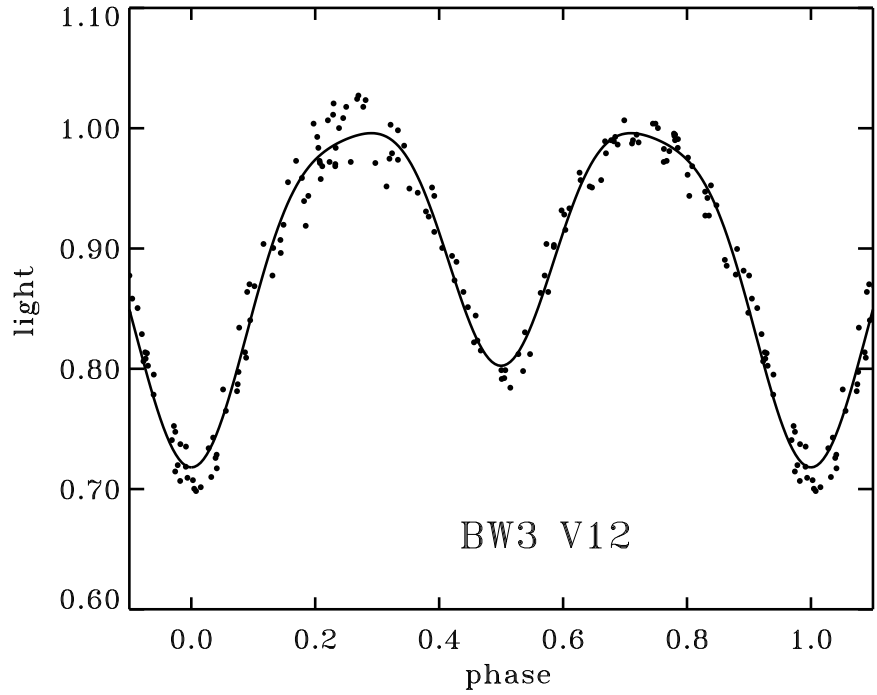,height=3.5in}}
\vskip 0.5in
\caption{The light curve for \#3.012 which is presently the
shortest-period  
(0.370 day) known contact binary showing a moderately large
eclipse-depth difference. The line gives the fit based on the
first  five cosine terms (zero to four). Note the scatter of points in
the first maximum which might indicate a transient asymmetry, similar
to those frequently observed among systems showing
differences in depths of eclipses (see the next section).}
\end{figure}

\begin{figure}           
\centerline{\psfig{figure=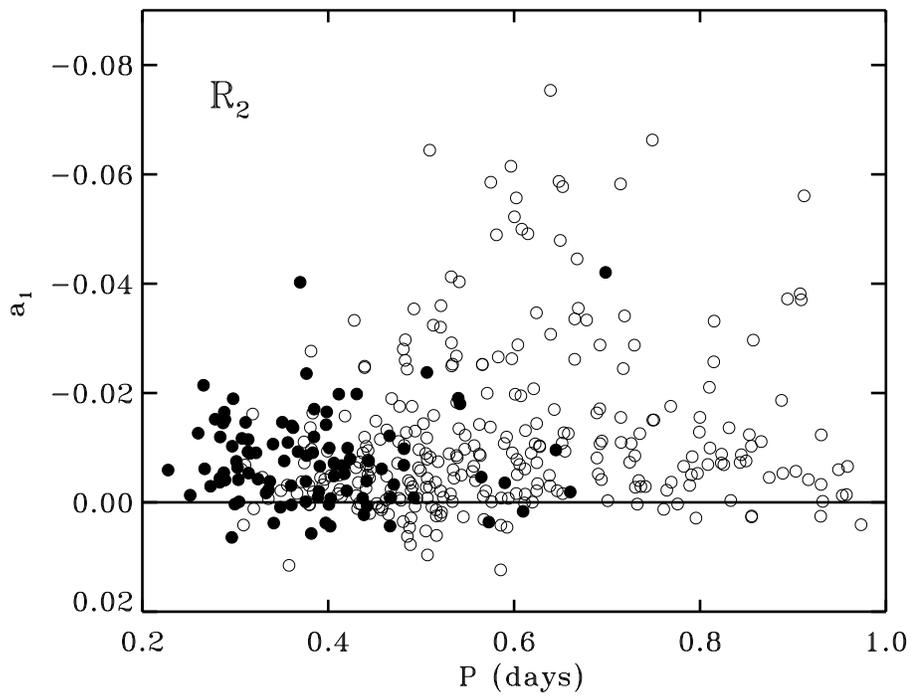,height=3.5in}}
\vskip 0.5in
\caption{The $a_1$ coefficients plotted versus the orbital period in
the same format as in Figure~3. Note the increase in frequency
of occurrence of  unequally-deep minima  among systems
with longer periods. The two local systems are \#3.012 at 0.370
day and \#6.005 at 0.698 day.}
\end{figure}

\begin{figure}           
\centerline{\psfig{figure=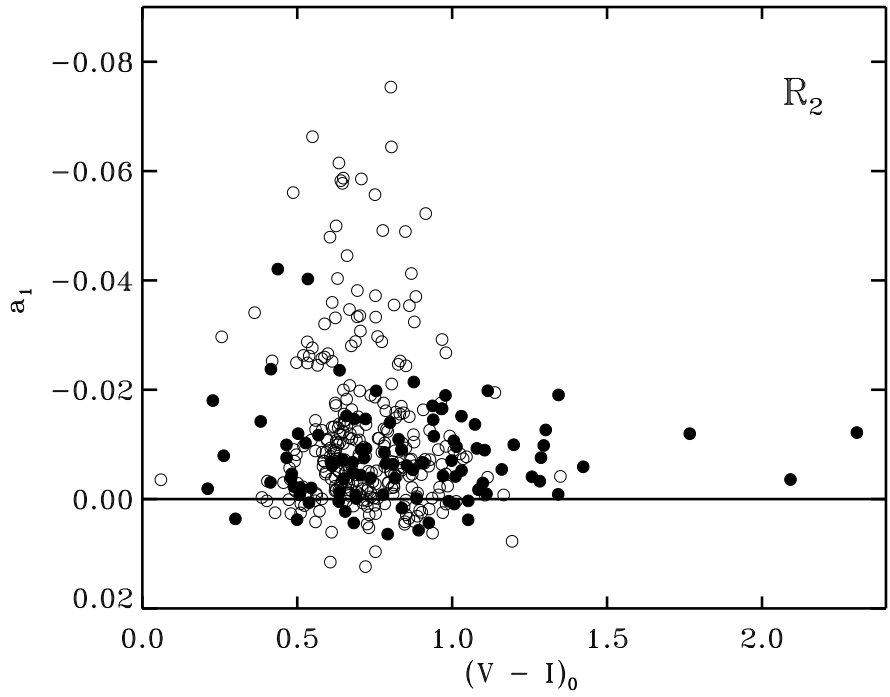,height=3.5in}}
\vskip 0.5in
\caption{The same data as in Figure~5 
plotted versus the intrinsic color. The
three very red systems are described in the text.}
\end{figure}

\begin{figure}           
\centerline{\psfig{figure=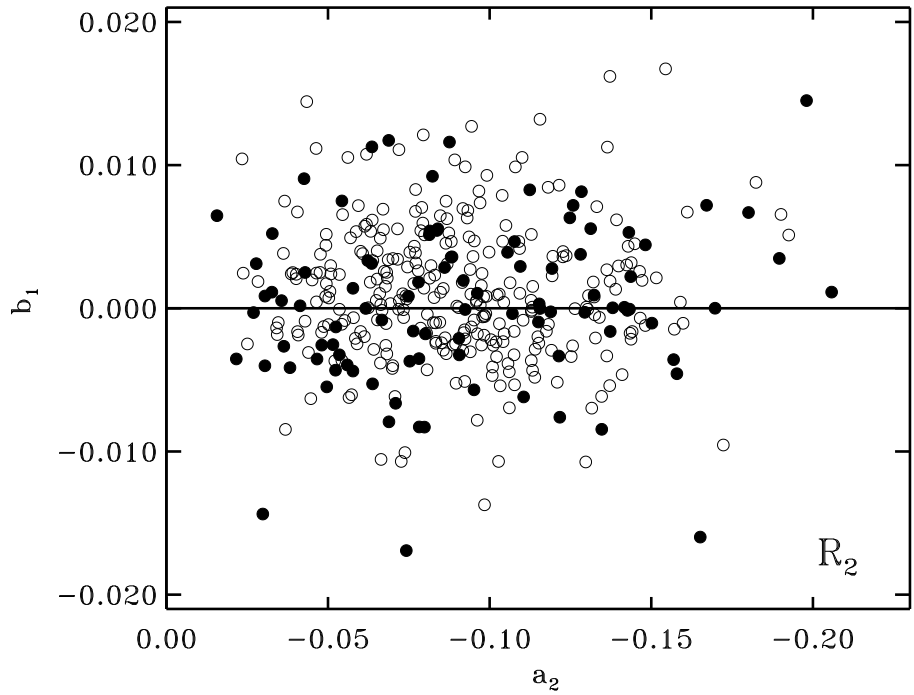,height=3.5in}}
\vskip 0.5in
\caption{The first sine coefficient in the Fourier representation of
the light
curves, $b_1$, is shown here versus the largest even coefficient
$a_2$. Although the diagram seems to indicate a random scatter
in $b_1$ without any sign preference, the positive values of $b_1$
are more frequent than the negative values for intrinsically-bright
and more distant systems with long orbital periods; see the next
figure.}
\end{figure}

\begin{figure}           
\centerline{\psfig{figure=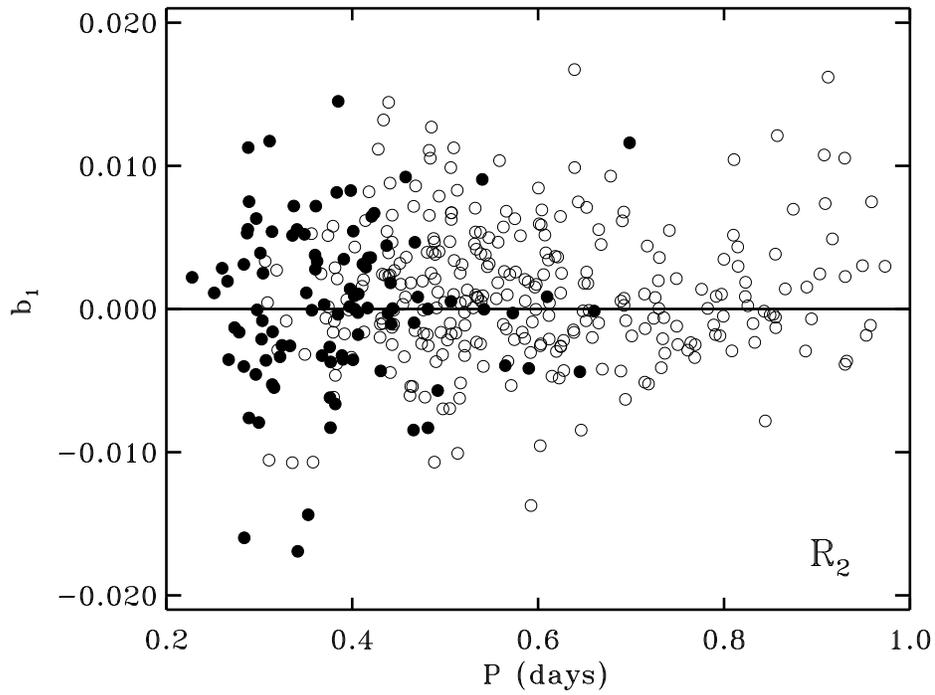,height=3.5in}}
\vskip 0.5in
\caption{Positive values of the sine coefficients $b_1$ are
systematically  more
common among contact systems with orbital periods longer than
about 0.4 days. The local-sample systems, which in their majority
have periods shorter than 0.4 day, seem to show a symmetric
distribution of the $b_1$ values.}
\end{figure}

\begin{figure}           
\centerline{\psfig{figure=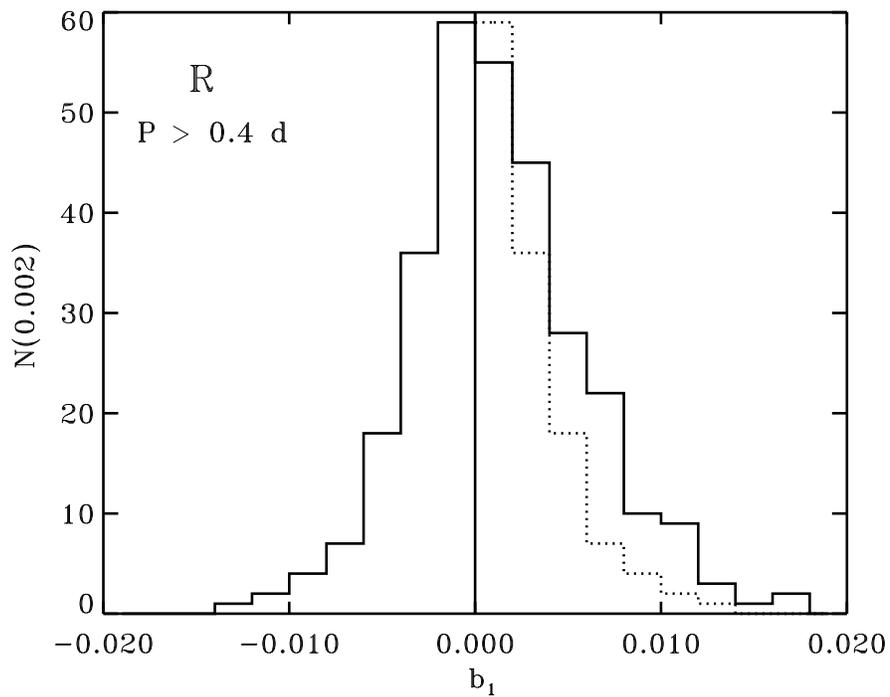,height=3.5in}}
\vskip 0.5in
\caption{The histogram of $b_1$ for all systems in the R-sample with
the
orbital periods longer than 0.4 day. The dotted line visualizes
the difference of the two sides of the distribution
by giving the reflection of the negative part of
the histogram into the positive side.}
\end{figure}

\begin{figure}           
\centerline{\psfig{figure=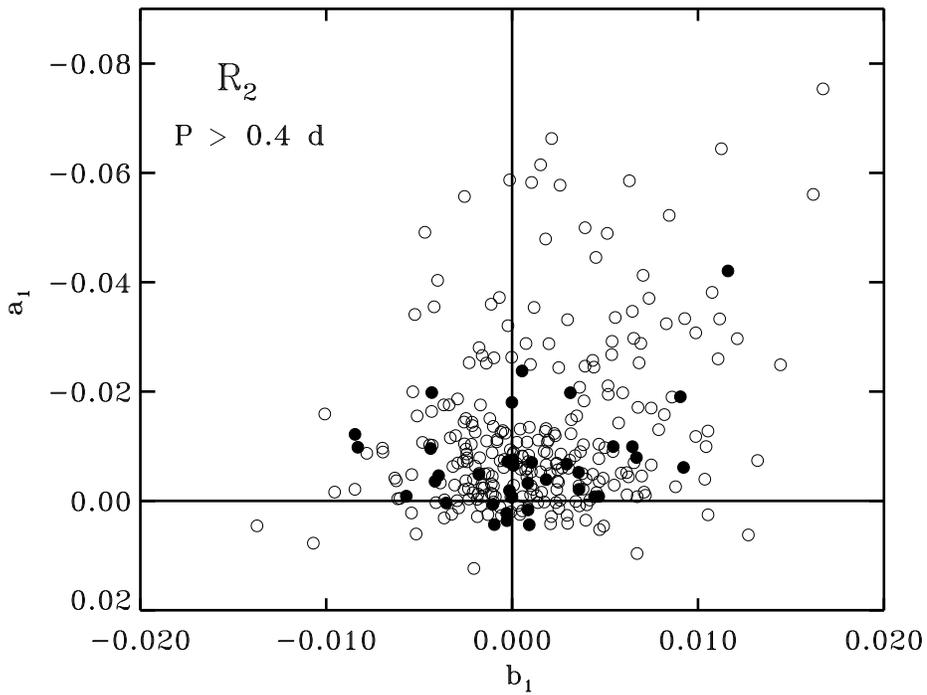,height=3.5in}}
\vskip 0.5in
\caption{Relation between $b_1$ and $a_1$ coefficients for the
R-sample systems
with periods longer than 0.4 day. $b_1$ measures the
light-curve asymmetry and $a_1$ measures the difference in depth of
eclipses. The correlation between these quantities
has a 3-sigma level significance. Note the single filled circle in the
upper right quadrant which corresponds to the system \#6.005;
its light curve is shown in the next figure.}
\end{figure}

\begin{figure}           
\centerline{\psfig{figure=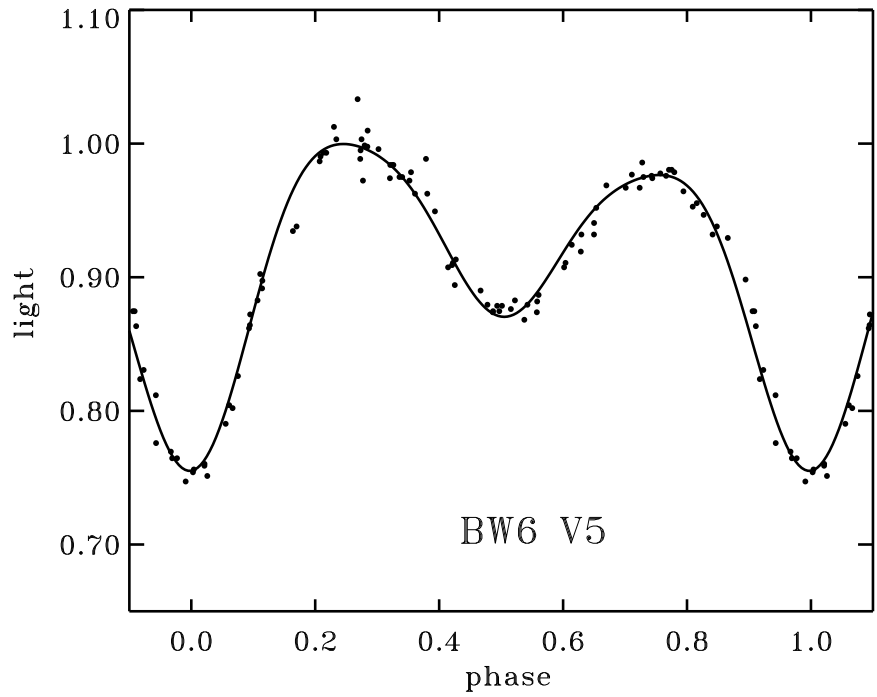,height=3.5in}}
\vskip 0.5in
\caption{The light curve for the binary \#6.005 (orbital period
0.698 day) which shows the largest
asymmetry of maxima among the systems of the local sample. This system
shows also a large difference in the depth of the eclipses. Note the
sense of the asymmetry which is apparently the most common among
similar systems. The fit, shown by a continuous line,
consists of 5 cosine terms (zero to four) and one sine term.}
\end{figure}

\begin{figure}           
\centerline{\psfig{figure=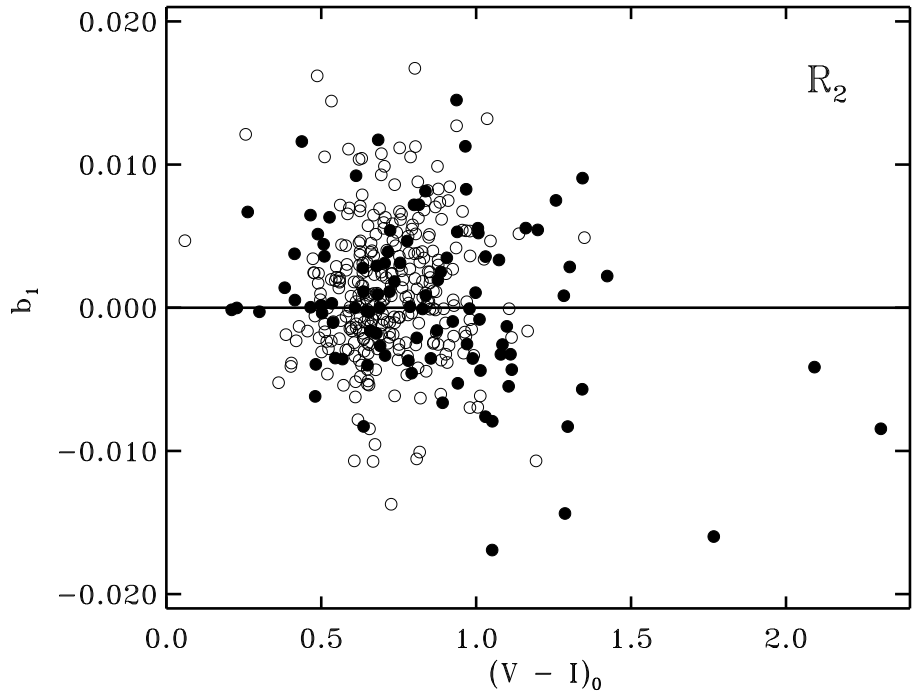,height=3.5in}}
\vskip 0.5in
\caption{Relation between $b_1$ and the intrinsic colors. Note that
all of the few very red systems of the local sample 
show  negative values of $b_1$.} 
\end{figure}

\begin{figure}           
\centerline{\psfig{figure=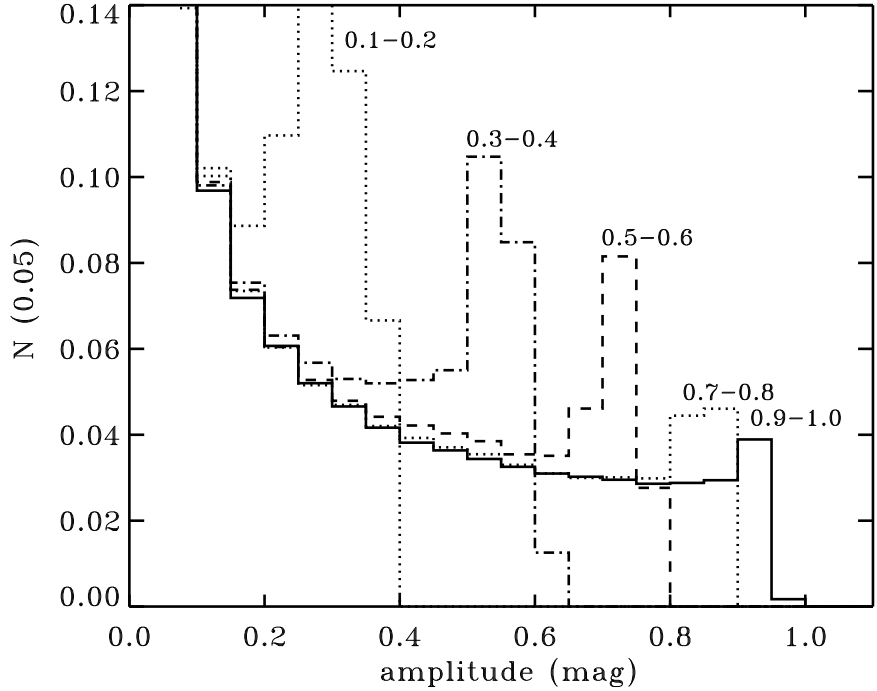,height=3.5in}}
\vskip 0.5in
\caption{The mass-ratio distribution is expected to strongly
influence the distribution of the variability amplitudes. Here, the
 theoretical distributions, based on the data in R93,
 are shown for idealized $Q(q)$ distributions
consisting of flat segments within $\Delta q = 0.1$ ranges
of $q$, as labeled in the figure. For a broad $Q(q)$, the broad spikes
due to higher frequency of total eclipses for narrow ranges in $q$
become diluted in the combined amplitude distribution. The low
amplitude end of the distribution 
is practically insensitive to changes in $Q(q)$.}
\end{figure}

\begin{figure}           
\centerline{\psfig{figure=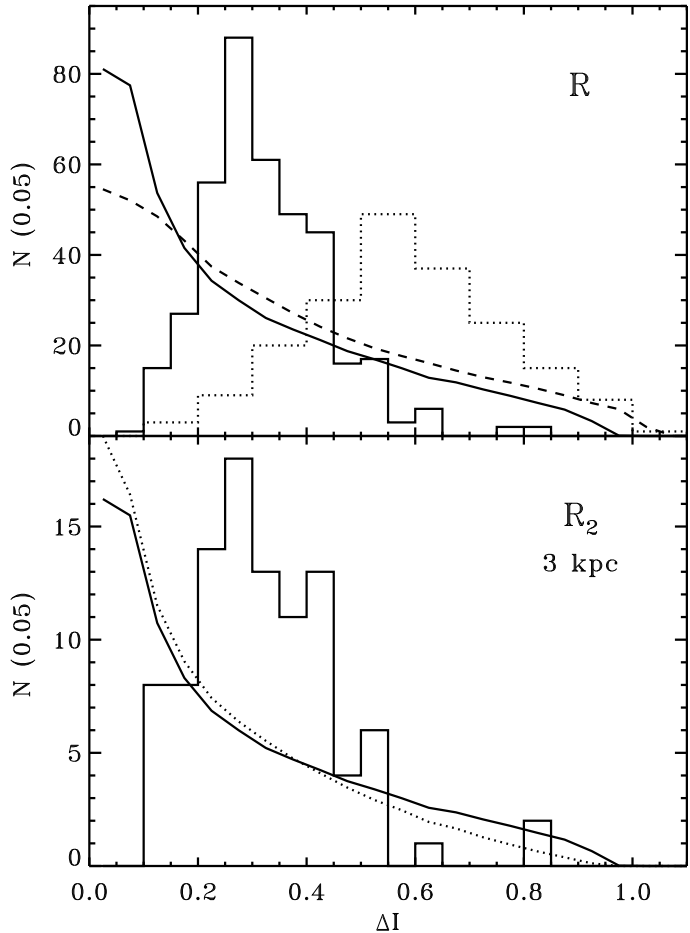,height=4.5in}}
\vskip 0.5in
\caption{The observed amplitude distributions for the R-sample
(histograms) are compared here with the general field
distribution (dotted histogram)
and with some theoretical predictions (lines). The upper panel gives
the histogram for the whole R-sample while the lower panel is for the
local sub-sample to 3 kpc. The theoretical predictions are
shown by lines with an arbitrary normalization to the
total number of 100 systems. For both panels,
the same continuous line gives the expected amplitude
distribution for $Q(q) = const$ and for the inner contact,
$f=0$. Additionally, the broken line in the upper panel gives the case
$Q(q) = const$ and $f=0.5$, whereas the dotted line
in the lower panel gives the inner contact case, $f=0$,
but for a falling mass-ratio distribution, $Q(q) = 1-q$.
Note the weak sensitivity of the theoretical
distributions to these modifications. The theoretical distributions
appear to be very different 
the observed distributions which steeply rise for amplitudes below
$\Delta I < 0.6$; unfortunately, the latter are affected by
the detection incompleteness for $\Delta I < 0.3$ (R97).
The sample of the sky-field systems
(dotted histogram) is heavily biased by the ease of detection of large
amplitude systems which do exist but are
apparently very rare (see the text).}
\end{figure}

\end{document}